\newif\ifFull

\Fullfalse

\ifFull
\documentclass{article}
\else
\documentclass{llncs}
\fi

\usepackage{fullpage}
\usepackage {amsmath}
\usepackage {amssymb}
\usepackage{color}
\usepackage {plaatjes}
\usepackage {xspace}
\usepackage {cite}
\usepackage {url}

\definecolor {namecolor} {rgb} {0.9,0.1,0.3}
\definecolor {remarkcolor} {rgb} {0.1,0.7,0.2}

\let\oldendproof\endproof
\def\endproof{\qed\oldendproof}

\ifFull
\newtheorem {theorem} {Theorem}
\newtheorem {lemma} {Lemma}
\newtheorem {definition} {Definition}
\newenvironment {proof} {\textbf {Proof:}} {\hfill \ensuremath {\boxtimes}}
\fi

\newcommand{\routing}{\textsf{ROUTING}\xspace}
\DeclareMathOperator{\diam}{diam}
\DeclareMathOperator{\md}{memdim}
\DeclareMathOperator{\member}{cat}
\DeclareMathOperator{\dist}{sp}
\DeclareMathOperator{\depth}{depth}
\DeclareMathOperator{\height}{height}
\DeclareMathOperator{\ancestors}{ancestors}
\DeclareMathOperator{\leftrm}{left}
\DeclareMathOperator{\rightrm}{right}

\title{Category-Based Routing in Social Networks:\\ 
    Membership Dimension and
   the Small-World Phenomenon}
\author{David Eppstein \and Michael T. Goodrich \and Maarten L\"offler \and
Darren Strash \and Lowell Trott}
\institute{Dept. of Computer Science, University of California, Irvine, USA}

\begin {document}

\pagestyle{plain}

\maketitle

\begin{abstract}
A classic experiment by
Milgram shows that individuals can route messages along short paths
in social networks,
given only simple categorical information about recipients (such
as ``he is a prominent lawyer in Boston'' or ``she is a Freshman sociology
major at Harvard'').
That is, these networks have very short paths between pairs of nodes (the so-called \emph{small-world phenomenon});
moreover, participants are able to route messages along these paths
even though each person is only aware of a small part of the 
network topology. Some sociologists conjecture that participants in such scenarios use a \emph{greedy routing} strategy in which they forward messages to acquaintances that have more categories in common with
the recipient than they do, and similar strategies have recently been proposed for routing messages in dynamic ad-hoc networks of mobile devices.
In this paper, we introduce a network property called
\emph{membership dimension}, 
which characterizes the cognitive 
load required to maintain relationships between participants and
categories in a social network.
We show that any connected network has a system of categories that will support greedy routing, but that these categories can be made to have small membership dimension if and only if the underlying network exhibits the small-world phenomenon.
\end{abstract}


\section{Introduction}

In a pioneering experiment in the 1960's, Stanley Milgram and 
colleagues~\cite{milgram67,travers-milgram69,korte-milgram70}
empirically studied the ability of people in real-world social networks
to route messages to their acquaintances, and used their studies to deduce properties of these networks.
296 randomly chosen individuals in Omaha,
Nebraska and Wichita,
Kansas were asked to forward a letter to a lawyer in 
Boston by using the following rule:
send the letter to an acquaintance so that it progresses toward the recipient.
Each acquaintance along the way
is then told to forward the letter by this same rule.
The results of these experiments reveal that,
if a message gets to its recipient,
it typically passes between at most six
acquaintances\footnote{This observation has also led to the 
    concept of ``six degrees of separation''
    between all people on earth
    and the trivia game, ``Six Degrees of Kevin Bacon,'' where players
    take turns trying to link performers
    to the actor Kevin Bacon
    via at most six movie collaborations.}---and this 
observation has come to be called the 
\emph{small-world phenomenon}~\cite{k-swpa-00,w-ndswp-99}.

What is perhaps even more surprising than the existence of these short 
paths is the fact that human participants are able to 
efficiently route messages using only local information and 
simple facts about message targets, such as 
gender, ethnicity, occupation, name, and location. 



As a way to study how humans can route such messages, 
several groups of sociology researchers have studied 
the importance of \emph{categories},
that is, various groups to which people belong,
in the small-world phenomenon.
For instance,
in the early 1970's,
Hunter and Shotland~\cite{hs-tdcsw-74} found that messages routed
between participants who both belonged to the same  category of people in a university (such as students, faculty, or
administrators) had shorter paths than messages routed across such
categories.
Along these same lines,
Killworth and Bernard~\cite{Killworth-Reverse} 
performed a set of experiments in the late 1970's they called 
\emph{reverse small-world experiments}.
In these experiments, they presented each participant with a list of
messages for hundreds of targets, identified by the categories of 
town, occupation, ethnic background, and gender,
and they asked each such participant to whom they would
send each of these messages.
One of the main conclusions of this study was that the choices people
make in deciding on routes are overwhelmingly categorical in nature.
In the late 1980's, Bernard {\it et al.}~\cite{bkems-ssrcc-88}
extended this work to identify which of twenty categories are the most
important to people from various cultures for the sake of message routing.
More recently,
Watts {\it et al.}~\cite{Watts02identityand} 
present a hierarchical model for categorical organization in social 
networks for the sake of message routing. They propose groups
as the leaves of rooted trees, with internal nodes defining
groups-of-groups, and so on. They define an \emph{ultrametric} on the vertices of each hierarchy (a distance function in which the distance between any two participants is determined by the level in the hierarchy of the smallest category containing both of them)  and they conjecture that people use the minimum distance in one of their trees to make message routing 
decisions.
That is, they argue that
individuals can understand their ``social distance" to a target
as the minimum of the distances between them and the target in each of their
hierarchical categories.
Of course, such a determination requires some global knowledge about
the structures of the various group hierarchies.

Although this previous work shows the importance of categories and of hierarchies of categories in explaining the small world phenomenon, it does not explain where the categories come from or what properties they need to have in order to allow greedy routing to work.  
Hence, this prior work leaves open the following questions:
\begin{itemize}
\item
Which social networks support systems of categories that allow participants to route messages using the simple greedy rule of sending
a message to an acquaintance who has more categories in common with
the target? 
\item
How complicated a system of categories is needed for this purpose, how much information about this system do individual participants need to know, and what properties of the underlying network can be used to characterize the complexity of the category system?
\end{itemize}



Our goal in this paper, therefore, is to address these questions
by studying the existence
of mathematical and algorithmic 
frameworks that demonstrate the feasibility of local, greedy, category-based
routing in social networks.

\subsection{Our Results}

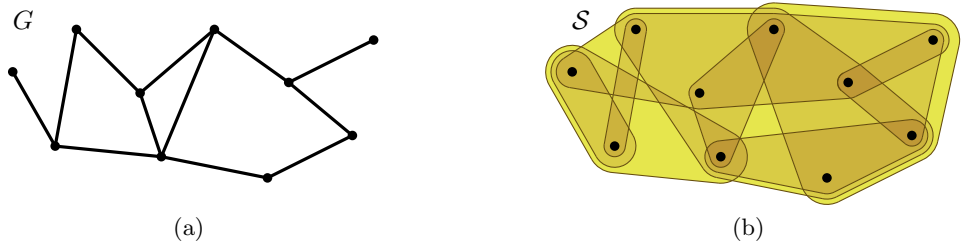
\tweeplaatjes {ex-graph} {ex-groups} {A set of elements $U$ (drawn arbitrarily as points in the plane). (a) The graph $G$ on $U$. (b) The categories $\mathcal S$ on $U$. In this example, the membership dimension is $4$, because no element is contained in more than $4$ groups.}

Inspired by the work of
Watts {\it et al.}~\cite{Watts02identityand},
we view a social network as an undirected graph $G=(U,E)$, whose
vertices represent people and whose edges 
represent relationships, taken together with a collection,
${\mathcal S} \subset 2^{U}$, of categories defined on the vertices
in $G$. Although the categories that we end up constructing in proving our results will have a natural hierarchical structure, we do not impose such a structure as part of our definitions.
Figure~\ref {fig:ex-graph+ex-groups} shows an example.

In addition, 
given such a social network, $G=(U,E)$ with a category system ${\mathcal S}$,
we define the \emph{membership dimension}
of $\mathcal S$ to be 
\[
\max_{u\in U} |\{ C\in {\mathcal S}\colon\, u\in C\}|,
\]
that is,
the maximum number of groups to which any one person in the network 
belongs.  The membership dimension characterizes the \emph{cognitive load} of performing routing tasks in the given system of 
categories---if the membership dimension is small, each actor in the network only needs to know a proportionately small amount of information about his or her own categories, 
his or her neighbors' categories,
and the categories of each message's eventual destination.
Thus, we would expect real-world social networks to have small
membership dimension.

In this paper, we provide a constructive proof that a category system with low membership dimension can support greedy routing. Our results are not intended to model the actual formation of social categories, and we take no position on whether categories are formed from the network, the network is formed from categories, or both form together. Rather, our intention is to show the close relation between two natural parameters of a social network, its path length and its membership dimension.
In particular:
\begin{itemize}
\item
We show that the membership dimension 
of $(G,{\mathcal S})$
must be at least the diameter of $G$, $\diam(G)$, 
for a local, greedy, category-based routing strategy to work.
\item
Given any connected graph $G=(U,E)$, 
we show there is a collection, $\mathcal S$, of categories
defined on the set, $U$, of vertices of $G$, 
such that local, greedy, category-based routing always works.
Moreover, the
membership dimension of $(G,{\mathcal S})$ in this case 
is $O((\diam(G)+\log |U|)^2)$.
\item We show that some dependence between the category system and the underlying graph is essential, by proving that there does not exist a single category system that supports greedy routing  regardless of its underlying graph.
\end{itemize}
Since the earliest work of
Milgram~\cite{milgram67,travers-milgram69,korte-milgram70},
social scientists have generally believed the so-called ``small world hypothesis'' that the diameters of real-world social
networks are bounded by small constants or by slowly growing functions of the network size.
Under a weak form of this assumption, that the diameter is $O(\log |U|)$,
our results provide a natural model for how participants in a social
network could efficiently route messages 
using a local, greedy, category-based routing strategy
while remembering an amount of
information that is only polylogarithmic in the size of the network.

%

\subsection{Previous Related Work}

\paragraph{Greedy Routing.}
In addition to the greedy method described by 
Milgram~\cite{milgram67,travers-milgram69,korte-milgram70}
for routing in social networks,
geometric greedy routing~\cite{Couto01locationproxies,Kranakis99compassrouting}
has been introduced in computer communications
as a method to leverage the geographic location of nodes in 
ad-hoc and sensor networks in order to reduce the computational overhead of routing 
messages.  
In geometric greedy routing, vertices have coordinates 
in a geometric metric space. They use these coordinates to calculate the distances between the destinations of a message and their neighboring vertices; each message is routed greedily, to a neighbor that is closer to the message's destination.
Not every geographic network has the property that this strategy will correctly route all messages to their destinations, so a number of
techniques have been developed to assist such greedy routing schemes
when they fail~\cite{bose-GaurDel-01,Karp:2000:GGP:345910.345953,Kuhn:2002:AOG:570810.570814,Kuhn:2003:WOA:778415.778447,Kuhn:2003:GAR:872035.872044}.
In a paradigm introduced by Rao et al.~\cite{Rao:2003:GRW:938985.938996},
virtual coordinates can also be introduced to overcome the shortcomings of 
real-world coordinates and allow simple greedy forwarding to function 
without the assistance of fallback algorithms. This approach
has been explored by several other 
researchers~\cite{Papadimitriou:2005:CRG:1121826.1121828,leighmoit-someresul-10,Angelini-GDTriangulations-09,Klein-Hyp-07},
who study various network properties that allow coordinates to be found that will cause greedy routing to succeed.
In addition, several researchers also study the existence of 
\emph{succinct} virtual coordinate systems~\cite{Muh-dis-07,Maymounkov06greedyembeddings,Eppstein:2009:SGG:1506879.1506884,Goodrich-SuccEuclid-09},
where the number of bits needed to represent the coordinates of each vertex is polylogarithmic in the
size of the network. If an assignment of virtual coordinates is succinct in this way, then the amount of computer memory needed to store the coordinates will be significantly smaller than the memory needed for a complete routing table that avoids the need for greedy routing. This notion of succinctness, and its motivation in reducing memory requirements, is closely analogous to our definition of the membership dimension for categorical greedy routing.  Just as succinct bit representation is required to make greedy routing space-efficient, sociological routing requires low membership dimension to reduce the cognitive load on its participants and make it feasible for them to participate.

Almost all of this
previous work on greedy routing in computer networks uses vertex coordinates in
$2$- and $3$-dimensional Euclidean or hyperbolic spaces.
However, one very recent exception to this restriction
is work by Mei {\it et al.}~\cite{Mei-Cat-11},
who study category-based greedy routing as a heuristic
for performing routing in dynamic delay-tolerant networks of computing devices. Mei {\it et al.} assume that
the network nodes have been organized into pre-defined categories based on
their owners' interests. Their
experiments suggest that using these categories for greedy routing is superior in practice to
routing heuristics based on location or simple random choices.
It is possible to interpret the categorical greedy routing techniques of Mei {\it et al.} and of this paper
as being geometric routing schemes using virtual coordinates, where the coordinates of each node represent their category memberships. In this interpretation, the membership dimension of an embedding corresponds to the number of nonzero coordinates of each node, and our results
show that such greedy routing schemes can be done succinctly in
graphs with small diameter.

\paragraph{The Small-World Phenomenon Through an Algorithmic Lens.}
Like the work of this paper,
Kleinberg~\cite{k-swpa-00} studies the 
small-world phenomenon from an algorithmic perspective.
His approach takes an orthogonal direction from our work, however, in two ways.
First, he focuses exclusively on location as the critical factor for
supporting the small-world phenomenon (under a geometric metric), whereas
our work focuses on greedy routing strategies based on 
categories and membership dimension.
Second, his study takes vertex coordinates as a given and constructs
the network from these coordinates based on geometry and random choices, whereas
our approach takes the network as a given and studies the kinds of
categorical structures needed to support category-based greedy
routing.

In addition to this work by Kleinberg, 
many other researchers have proposed various different
models for randomly generating graphs that possess properties similar to 
those in real-world social networks, such as being scale-free (obeying a power law in the degree distribution) or having small diameter.
For instance, see~\cite{Duchon200696,RSA10084,mn-akswm-04,n-msw-00,ws-cdswn-98}.



\section {Routing in Networks based on Categorical Information}

In this section, 
we introduce a mathematical model of categorical greedy routing.
This model defines in a precise way the routing strategy that we hypothesize 
people use to route messages in a real-world social networks,
based on prior 
work~\cite{bkems-ssrcc-88,hs-tdcsw-74,Killworth-Reverse,Watts02identityand}.
Additionally, we provide some basic definitions and properties that, when they hold for a network, allow us to guarantee the success of this routing strategy.
  
  \subsection {Basic definitions}
  
    Abstracting away the social context, let $U$ be the universe of
    $n$ people defining the potential sources, targets, and intermediates for
    message routes, and
    let $G=(U,E)$ be an undirected graph on $U$ whose $m$ edges represent 
    pairs of people who can send messages to each other. 
    
    \begin{definition}[diameter]
    For any two elements $s, t \in U$, we define $\dist(s,t)$ to be the length of the shortest path between $s$ and $t$ in $G$. Then the diameter of $G$, denoted $\diam(G)$, is $\max_{s,t \in U} \dist(s,t)$, the maximum length of any shortest path in $G$.  That is, it is the distance between the two vertices that are farthest from each other in $G$.
    \end{definition}
    
    In the greedy routing algorithms that we study, a central concept is a \emph{neighborhood}, the set of participants that a message could be forwarded to.
   \begin{definition}[neighborhood]
    For $s \in U$, 
    we define the \emph {neighborhood}, $N(s)$,
    to be the set of neighbors of $s$ in $G$, 
    that is,
    $$N(s) = \{u \in U \mid \{s,u\} \in E\}.$$
    \end{definition}
    
    Moving from graphs to category systems, we define the membership dimension, a numerical measure of the complexity of a system of categories that is fundamental to our work.
    
    \begin {definition}[membership dimension]
Let ${\mathcal S} \subset 2^{U}$ be a set of subsets of $U$, which represent the abstract categories that elements of $U$ can belong to.
    For a given $u \in U$, we define $\member(u) \subset \mathcal S$ to be the set of groups to which $u$ belongs:
    $$\member(u) = \{C \in \mathcal S \mid u \in C\}.$$
          The \emph {membership dimension} of $\mathcal S$ is the maximum number of elements of $\mathcal S$ that any element of $U$ is contained in, that is,
      \[
      \md(\mathcal S) = \max_{u\in U} |\member (u)|.
      \]
    \end {definition}

    As discussed in the introduction, there is reason to believe that in real world social networks and group structures $(G,\mathcal S)$, both $\diam (G)$ and $\md(\mathcal S)$ tend to be small.


  \subsection {The routing strategy}
  
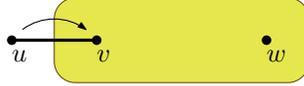
    \eenplaatje {routing} {Illustration of the routing rule. $v$ is a viable candidate for forwarding from $u$ because $v$ and $w$ share more category memberships than $u$ and $w$.}

    We now describe a simple category-based strategy to route a message from some node $s \in U$ to another node $t \in U$.    The strategy is greedy, and therefore follows the \emph{greedy routing rule}.  We clarify the distance function following the definition:

    \begin {definition}[greedy routing rule]
      If a node $u$ receives a message $M$ intended for a destination $w \neq u$, then $u$ should forward $M$ to a neighbor $v \in N(u)$ that is closer to $w$ than $u$ is, that is, for which $d(v,w) < d(u,w)$.
    \end {definition}


    
    As mentioned above,
    the distance function we study is category-based, and measures the number of shared groups of $\mathcal S$ that two nodes belong to.
    In particular, we define the distance $d(s,t)$ by the formula
    $$d(s,t) = |\member(t) \setminus \member(s)|.$$
    The backslash denotes the set-theoretic difference operator,
    so this distance function\footnote{Note that $d$ might not determine a metric space, because it need not necessarily be symmetric.}
    measures the number of categories of the target that the current node does \emph {not} share. This number decreases as the number of shared groups of $\mathcal S$ between the current node and the target increases.
    Figure~\ref {fig:routing} illustrates the routing rule.
    We refer to the greedy routing strategy that uses this distance function as \routing.

    In real-world networks for which $\md (\mathcal S)$ is small (as we conjecture), this strategy 
    should 
    be easy for participants to perform.  A small $\md (\mathcal S)$ makes it feasible for each participant to be aware of the 
    categories to which he himself, his neighbors, and the target belong, and therefore allows the participants in the network to perform greedy routing with only a small cognitive load.


  \subsection {Successful routing}
  
    We now investigate under what conditions \routing can be successful in routing a message between any pair of nodes in a network. We identify several properties of a graph $G$ and associated group structure $\mathcal S$ 
    that directly influence the feasibility of the routing strategy.
    
    For routing to be possible, $G$ must be connected.
    But it seems natural to consider a stronger property, \emph{internal connectivity}, which we define below.

    \begin{definition}[restriction]
      If $G$ is a graph, $\mathcal S$ is a category system for $G$, and $C$ is a category in $\mathcal S$, then the \emph{restriction} of $G$ to $C$ is the subgraph of $G$ induced by $C$. That is, it is the graph with $C$ as its vertex set and with an edge connecting every two vertices in $C$ that are adjacent in~$G$.
      \end{definition}

    \begin {definition}[internal connectivity]
      A pair $(G, \mathcal S)$ is  
      \emph{internally connected}
      if for each $C \in \mathcal S$, $G$ restricted to $C$ is connected.
    \end {definition}

    Figure~\ref {fig:ic-not-sh} shows an example of an internally connected pair.
    This is a very natural property for sociological groups to exhibit. 
    People belonging to the same group will have greater cohesiveness, 
    and if a group fails the condition to be \emph {internally connected},
    then the group can be redefined sensibly to be the 
    set of groups defined by their connected components.

    \tweeplaatjes {ic-not-sh} {sh-not-ic} 
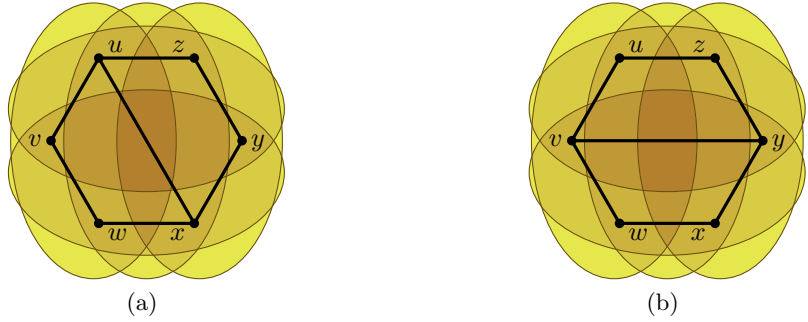
    {Two examples with the same set of elements $U = \{u,v,w,x,y,z\}$ and categories $\mathcal S = \{\{u,v,w\},\{x,y,z\},\{u,w,x,z\},\{u,v,y,z\},\{v,w,x,y\}\}$.
    (a) An example that is internally connected, but not shattered: there is no neighbor of $v$ that shares a region with $y$ that $v$ is not in. (b) An example that is shattered, but not internally connected: the induced graph of $\{u,w,x,z\}$ is not connected.}
  
    \begin {definition}[shattered]
      A pair $(G, \mathcal S)$ is \emph {shattered} if,
      for all $s, t \in U$, $s \neq t$, there are a neighbor $u \in N(s)$
      and a set $C \in \mathcal S$ such that $C$ contains $u$ and $t$, but not $s$.
    \end {definition}

    Figure~\ref {fig:sh-not-ic} shows an example of a shattered pair.
    Note that in this definition, $u$ and $t$ could be the same node.
    This property falls out naturally from the instructions given in the
    real-world routing experiments of Milgram and others. 
    In order for someone to advance a letter toward a target, 
    there must be an acquaintance that shares additional interests 
    with the target. 
    Indeed, we now show that the shattered property is 
    necessary for \routing to work.

\begin{lemma} \label {lem:shattered-necessary}
If $(G, \mathcal{S})$ is not shattered, then \routing does not correctly route messages between all pairs of vertices.
\end{lemma}

\begin{proof}
Suppose that $(G, \mathcal{S})$ is not shattered. That is, there exists a pair of vertices $s$ and $t$, such that each category $C$ that is shared by $t$ and a neighbor of $s$ is also shared by $s$.
If this is the case, then it is not possible for any neighbor $u$ of $s$ to share strictly more categories with $t$ as $s$ does. Therefore, \routing will fail
to route messages from $s$ to $t$.
\end{proof}

    Furthermore, if $G$ is a tree, then these two properties of being shattered and of internal connectivity together are in fact \emph {sufficient} for the routing strategy to always work.

  \begin {lemma}
\label{lemma:treeroutingworks}
    If $G$ is a tree, and $(G,\mathcal{S})$ is internally connected and shattered, then \routing is guaranteed to route messages correctly between every pair of vertices.
  \end {lemma}

\begin{proof}
Let $s$ and $t$ be any two vertices in $G$. Since $G$ is a tree,
there is one simple path from $s$ to $t$. 
Let $(u,v)$ be an edge on the path from $s$ to $t$.

First, we claim that every category in $\mathcal S$ that contains both $u$ and
$t$ also contains $v$. This follows from the assumption that $(G, \mathcal S)$ is internally 
connected: any set $C \in \mathcal S$ with $u, t \in C$ must also contain $v$, 
since $v$ is on the only path between $u$ and $t$. 
Therefore, $v$ is contained in at least as many sets in $\mathcal S$
with $t$ as $u$ is.

However, by the assumption that $(G, \mathcal S)$ is shattered, $v$ must also share with $t$ a category in $\mathcal S$  that does not contain $u$.
Therefore $v$ shares strictly more categories with $t$ than $u$ does, so \routing will correctly forward a message addressed to $t$ from $s$ to~$v$.

Since we made no assumptions about $s$ and $t$ and showed that in each case \routing will always forward a message to the next vertex on a path to $t$, it follows that \routing succeeds for every pair of vertices.
\end{proof}

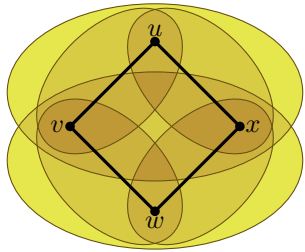
\eenplaatje {counter} {The \routing strategy does not work in this graph, even though it is internally connected and shattered. 
This graph has just four vertices, $U = \{u,v,w,x\}$, connected
in a cycle, taken together with the set of categories
${\mathcal S} = \{ \{u,v,x\},\, \{v,w,x\},\, 
       \{u,v\},\, \{v,w\},\, \{w,x\},\, \{u,x\}  \}$.
       However, \routing fails to route from $v$ to $x$, since
       $u$ is in 2 sets with $x$,
       $v$ is in 2 sets with $x$,
       and $w$ is in 2 sets with $x$.
       }

Although sufficient for routing in trees,
the internally connected and shattered properties are not sufficient 
for \routing to work on arbitrary connected graphs. 
Figure~\ref{fig:counter} shows a 
counter-example---\routing
is unable to route a message from the leftmost to the rightmost node, 
since there is no neighbor whose distance to the target is smaller.

\section {Existence of Categories}

In this section, we consider the following question: Is it possible to construct the family $\mathcal{S}$ so that
\routing always works and $\mathcal{S}$ has low membership dimension? 

We show that such a construction is always
possible if we are given a connected graph as input. We also show that it is impossible to construct
an $\mathcal{S}$ such that \routing will work if the graph is not known in advance. 

\subsection {Constructing $\mathcal S$ given $G$}

    Given a connected graph $G=(U,E)$ as input, we would like to construct a family $\mathcal{S}\subset 2^{U}$ so that
      \routing works, and the membership dimension of $S$ is small. 
      We concentrate foremost on constructions
    of category collections
    that are internally connected and shattered, 
    because of the social significance of these
    properties. 
    Nevertheless, even without these properties, we have the following lower bound.

    \begin {lemma}
      Let $G$ and $\mathcal S$ be a graph and a category system, respectively, such that  \routing works for $G$ and $\mathcal S$. Then $\md(\mathcal S) \geq \diam (G)$.
    \end {lemma}

    \begin {proof}
      Let $s$ and $t$, be any two vertices of $G$, and let $P$ be the path followed by \routing from $s$ to $t$. An edge $(u,v)$ can only be on $P$ if $d(v,t) < d(u,t)$.
      Since $d(\cdot,\cdot)$ can only take integer values, $d(u,t) \geq d(v,t) + 1$.
      It follows by induction on the length of $P$ that $d(s,t) \geq |P|$.
      
      Now, by the definition of the diameter of a graph, there exists a pair of vertices $s, t \in U$
      such that $sp(s, t) = \diam(G)$.
      Again, let $P$ be the path that \routing follows from $s$ to $t$; since the length of this path must be at least the length of a shortest path between the same two vertices, the length of $P$ is
      at least $\diam (G)$.
      
      By definition, $d(s,t) = |\member(t) \setminus \member(s)|$, and $\md(\mathcal S)$ is the maximum of $\member (\cdot)$ over all elements. Putting these definitions together with the inequalities we have deduced between $d(s,t)$, $|P|$, and $\diam(G)$, we have
      \[
        \md(\mathcal S) \geq |\member(t)| \geq |\member(t) \setminus \member(s)|
        = d(s,t) \geq |P| \geq \diam(G),
      \]
      as claimed.
    \end {proof}

    For paths, this bound is tight:

    \begin {lemma}
    \label{lem:path-is-shattered}
      If $G$ is a path, then there exists a category system $\mathcal S$ for $G$ such that $(G, \mathcal S)$ is shattered and internally connected and such that $\md(\mathcal S) = \diam(G)$.
    \end {lemma}

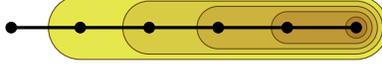
    \eenplaatje {paths} {The sets $B_v$ for each vertex $v$ in the path. The sets $A_v$ are constructed symmetrically. }

    \begin {proof} Arbitrarily pick one of the two end vertices of $G$ and 
    let us refer to the vertices in $G$ by their distance, $0$ to $n-1$, 
    from this vertex.
    For each vertex $i$, form two sets $A_i$ and $B_i$, 
    where $A_i=\{0,\ldots,i-1\}$ and $B_i=\{i+1,\ldots,n-1\}$, 
    and let $\mathcal{S} = \bigcup_{v\in U} \{A_v, B_v\}$.
    Figure~\ref{fig:paths} illustrates this construction.
    
    Each set in $\mathcal{S}$ consists of a path of vertices and therefore $\mathcal{S}$ is internally connected.
    $\mathcal{S}$ is also shattered, since for all $s$ and $t$, $s$ has a neighbor that shares either $A_s$ or $B_s$ with $t$, but $s$ is not in these sets.
    To calculate $\md(\mathcal{S})$, note that each vertex $i$ 
    is contained in sets $A_j$ for $0\leq j<i$ and $B_k$ for $k<i\leq n-1$.
    Therefore, each vertex is in exactly $n-1$ sets, which is $\diam(G)$. 
    \end {proof}

    A path is a special case of a tree. Therefore, whenever the given graph $G$ is a path, it follows from Lemma~\ref{lemma:treeroutingworks} and Lemma~\ref{lem:path-is-shattered}
 that it is possible to construct a category system $\mathcal{S}$ so that \routing works in  $G$ and so that the membership dimension 
$\md(\mathcal{S})$ equals $\diam(G)$, 

There are
also some other graphs, $G$, for which it is relatively easy 
to set up a category set, $\mathcal S$, that is 
shattered and internally connected in a way that supports the
\routing algorithm.
For example, in a tree of height~$1$ (i.e., a star graph),
with root $r$, we could simply create 
a separate category containing the root $r$
and each (leaf) child, plus a singleton category for each node.
Every path in this tree clearly supports the \routing strategy.
Note, however, that the
membership dimension of this category system is high, since the root
belongs to a linear number of categories. So even in this simple
example, supporting the \routing strategy and achieving 
a small membership dimension is a challenge.
Moreover, this challenge becomes even more difficult already for a
tree of height~$2$, since navigating from any leaf, $x$, to another leaf,
$y$, requires that the parent of $x$ belong to more categories with 
$y$ than $x$---and this must be true for \emph{every} other leaf,
$y$.
Thus,
it is perhaps somewhat surprising that we can construct a set of
categories, $\mathcal S$, for an arbitrary binary tree that causes
this network to be shattered and
internally connected (so the \routing strategy works, by
Lemma~\ref{lemma:treeroutingworks}) and such that $\mathcal S$
has small membership dimension.

    \begin{lemma} \label{lemma:binarytreememdim}
        If $G$ is a binary tree, 
  then there exists a category system $\mathcal S$ such that $(G, \mathcal S)$ is shattered and internally connected 
  and such that $\md(\mathcal S) = O(\diam^2(G))$.
    \end{lemma}

    \begin{proof} 
    We show how to construct $\mathcal{S}$ from $G$.
Arbitrarily pick a vertex $r\in U$ of degree at most $2$ and root the binary tree at $r$,
so each vertex $v$ has left and right children,
$\leftrm(v)$ and $\rightrm(v)$,
and let $\height(v)$ be the length of the longest simple path from $v$ to any descendant of $v$.
For each vertex $v$, we create a set $S_v$, containing $v$'s descendants (which includes $v$).
We further construct two families, $L_v$ and $R_v$,
using helper sets $L_{v,i}$ and $R_{v,i}$.
Let $L_{v,i}$ (resp., $R_{v_i}$) consist of $v$, 
the vertices in $v$'s left (right) subtree down to depth $i$,
and all vertices in $v$'s right (left) subtree.
Then define
\[
L_v = \{L_{v,i}\mid \depth(v)\leq i \leq \depth(v)+\height(\leftrm(v))\}.
\]
Figure~\ref {fig:bintree} illustrates this.
The family $R_v$ is defined symmetrically. Our $\mathcal{S}$ is then 
defined as
\[
{\mathcal S}\,  =\, \bigcup_{v\in U}{\{S_v\}\cup L_v \cup R_v}.
\]
By construction, each set in $\mathcal{S}$ is a connected subgraph of $G$ and therefore $\mathcal{S}$ is internally connected.
We can also see that $\mathcal{S}$ is shattered as follows.
If $s$ is an ancestor of $t$, then $s$'s child $u$ on the path to $t$ is contained in set $S_u$ which contains $u$, $t$, and not $s$.
Otherwise, let $v$ be the lowest common ancestor of $s$ and $t$, and assume without loss of generality that $s$ in $v$'s left subtree; then $s$'s parent is in $L_{v,\depth(s)-1}$ with $t$, and $s$ is not.

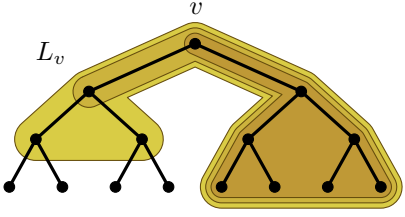
    \eenplaatje {bintree} {Showing the collection of sets $L_v$ for a small example subtree at $v$.}

We now analyze the membership dimension of this construction.
Let $v$ be a vertex, and let $\ancestors(v)$ be the set of $v$'s ancestors.
For $u\in \ancestors(v)$, $v\in S_u$, and $v$ belongs to $O(\height(u))$ sets of $L_u$ and $R_u$.
Then $v$ belongs to $O\left(\sum_{u\in \ancestors(v)}\height(u)\right)$ sets,
which is $O(\diam^2(G))$ for any $v$.
    \end{proof}

We now show how to extend this result to arbitrary trees. Our technique involves an application of \emph{weight balanced binary trees}\cite{knuth71,bent85}.

\begin{definition}[weight balanced binary tree]
A weight balanced binary tree is a binary tree
that stores weighted items in its leaves. If item $i$
has weight $w_i$, and all items have a combined weight of $W$ then item 
$i$ is stored at depth $O(\log {(W/w_i)})$. 
\end{definition}

\begin{lemma}
\label{lemma:embedtree}
Let $T$ be an $n$-node rooted tree with height $h$. We can embed $T$ into a binary tree such that the ancestor--descendant relationship is preserved, and the resulting tree has height $O(h + \log n)$.
\end{lemma}

\begin{proof}
Let $n_u$ be the number of descendants of vertex $u$ in $T$. For each vertex $u$ in $T$ that has more than two children, we expand the subtree consisting of $u$ and $u$'s children into a binary tree as follows. Construct a weight balanced binary tree $B$ on the children of $u$, where the weight of a child $v$ is $n_v$. We let $u$ be the root of $B$. Each child $v$ of $u$ in the original tree is then a leaf at depth $\log (n_u/n_v)$ in $B$. 
Performing this construction for each vertex $u$ in the tree expands $T$ into a binary tree with the ancestor--descendant relationship preserved from $T$.

Furthermore, each path from root to leaf in $T$ is only expanded 
by $\log (n)$ nodes, which we can see as follows.
Each parent-to-child edge $(u,v)$ in $T$ is replaced by a path of length 
$O(\log (n_u/n_v))$.
Therefore for each path $P$ from root $r$ to leaf $l$ in $T$, our construction expands $P$ by length $O(\sum_{(u,v)\in P}\log (n_u/n_v))$, 
which is a sum telescoping to $O(\log (n_r/n_l)) = O(\log n)$. 
Therefore, the height of the new binary tree is $O(h + \log n)$.
\end{proof}

Combining this lemma with Lemma~\ref{lemma:treeroutingworks}, we get the following theorem.

\begin{theorem}
\label{theorem:routetree}
Given a tree $T$, it is possible to construct a family $\mathcal{S}$ of subsets such that \routing works 
for $T$ and $\md(\mathcal{S}) = O((\diam(T) + \log n)^2)$.
\end{theorem}

\begin{proof}
Arbitrarily root $T$ and embed $T$ in a binary tree $B$ using the method in 
Lemma~\ref{lemma:embedtree}. Then $B$ has height $O(\diam(T) + \log n)$, and diameter $\diam(B) = O(\diam(T) + \log n)$. 
Applying the construction from Lemma~\ref{lemma:binarytreememdim}
to $B$ gives us a family $\mathcal{S}_B$
with $\md(\mathcal{S}_B)=O((\diam(T) + \log n)^2)$. 
We then construct a family $\mathcal{S}_T$, 
by removing vertices that are in $B$ but not $T$ from the sets in $\mathcal{S}_B$.
By construction, $(T,\mathcal{S}_T)$ is shattered and internally connected, and $\md(\mathcal{S}_T)\leq \md(\mathcal{S}_B)= O((\diam(T) + \log n)^2)$. By Lemma~\ref{lemma:treeroutingworks}, \routing works on $T$ with category
sets from $\mathcal{S}_T$.
\end{proof}

We can further extend this theorem to arbitrary connected graphs,
which is the main upper bound result of this paper.

\begin{theorem}
\label{theorem:memdimG}
      If $G$ is a connected graph, then there exists a category system $\mathcal S$ such that \routing correctly routes messages between all pairs of vertices and such that $\md(\mathcal S) = O ((\diam(G)+\log(n))^2)$.
\end{theorem}

\begin{proof}
Compute a low-diameter spanning tree $T$ of $G$. This step can easily be done using breadth-first search, producing a tree with diameter at most $2\diam(G)$. We then use the construction from Theorem~\ref{theorem:routetree} on $T$.
For greedy routing to work in a graph $G$,
note that
it is sufficient to show that it works in a spanning tree of $G$.
Therefore, since \routing works in $T$, \routing also works in $G$. 
\end{proof}

  \subsection{An Impossibility Result}
      It would be nice to construct a good group structure without knowing the structure of the graph in advance.
    Unfortunately, as we now show, this is impossible in general.

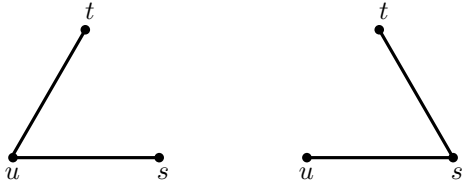
  \eenplaatje {impossible} {Two connected graphs on the same vertex set. Given the underlying vertex set, we cannot form a set of groups so that our greedy strategy routes from $s$ to $t$ in the left graph and from $u$ to $t$ in the right graph.}

\begin{theorem}
Given a set of vertices $U$, it is impossible to construct a set of groups $\mathcal{S}$ such that
our greedy routing strategy works on all connected graphs with $U$ as the vertex set.
\end{theorem}

\begin{proof}
Consider the two graphs in Figure~\ref{fig:impossible}. For \routing to route from $s$ to $t$ in
the left graph, $u$ must share more groups with $t$ than $s$ does. However, to route from $u$ to
$t$ in the right graph, $s$ must share more groups with $t$ than $u$ does. Both of these events
cannot happen simultaneously with one set of groups $\mathcal{S}$. Therefore \routing must fail in one
of these two graphs. 
\end{proof}
  
\section{Conclusion and Open Problems}

We have presented a construction of groups $S$ on a connected graph $G$
that allows a simple greedy routing algorithm, utilizing a
notion of distance on group membership, to guarantee delivery between nodes in 
$G$.
Such a construction will have membership dimension $O ((\diam(G)+ \log n)^2)$,
which demonstrates a reasonably small cognitive load for the members
of $G$.

There are several directions for future work.
For example,
while we have shown that the membership dimension must 
be minimally the diameter of $G$, it remains to be shown 
if the membership dimension must be the square of the diameter plus a
logarithmic factor for
arbitrary graphs.
We conjecture that the square term is not strictly needed in the
membership dimension in order for \routing to work.
Our group construction is performed for a general graph by selecting a low diameter spanning tree and using the presented tree construction,
so it may be possible that there is a group construction that 
has lower membership dimension and more efficient routing if 
it is constructed directly in $G$.

In addition, we observe that the groups that we
construct in our upper-bound proofs have a natural nesting property
that may correspond to a proximity-based way that people would organically
form groups.
It would be nice to verify or refute a hypothesis that people can
organize themselves in such groups using local information and simple
rules about how to form groups.

Finally,
we took the perspective in this paper that all categories have equal
weight with respect to routing tasks and that participants use
a simple greedy routing algorithm based solely on increasing the
number of categories in common with the target.
One possible direction for future work would be to define and study a
category-based routing strategy that allows participants to weight
various categories higher than others, as
in the work of Bernard {\it et al.}~\cite{bkems-ssrcc-88}.  This could include giving higher consideration to smaller or more well connected groups.  Another possible branch of further study might include analysis of the performance of this model when actors have only partial knowledge of the categories. A comparison could then be made between route lengths and level of category knowledge.



  \bibliographystyle {abbrv}
  \bibliography {refs,greedy}

\end {document}